\documentclass[aps, prd, twocolumn, nofootinbib, longbibliography,showkeys, superscriptaddress]{revtex4-1}

\usepackage{epsfig}
\usepackage[top=70pt,bottom=60pt,left=46pt,right=46pt]{geometry}
\usepackage{enumitem}
\usepackage{amsmath}
\usepackage{amssymb}
\usepackage{graphicx}
\usepackage{booktabs}
\usepackage{epstopdf}
\usepackage{gensymb}
\usepackage{aas_macros}
\usepackage{multirow}
\usepackage{lipsum}
\usepackage[english]{babel}

\usepackage[colorlinks=true,citecolor=blue,linkcolor=blue,urlcolor=blue]{hyperref}
\usepackage[capitalise,noabbrev]{cleveref}

\usepackage{tikz,xcolor,hyperref}
\DeclareUnicodeCharacter{2212}{-}

\definecolor{lime}{HTML}{A6CE39}
\DeclareRobustCommand{\orcidicon}{\hspace{-4pt}
	\begin{tikzpicture}
		\draw[lime, fill=lime] (0,0) 
		circle [radius=0.16] 
		node[white] {\hspace{0.1mm}{\fontfamily{qag}\selectfont \tiny ID}};
		\draw[white, fill=white] (-0.07,0.1) 
		circle [radius=0.01];
	\end{tikzpicture}
	\hspace{-3.2mm}
}
\foreach \x in {A, ..., Z}{\expandafter\xdef\csname 
	orcid\x\endcsname{\noexpand\href{https://orcid.org/\csname orcidauthor\x\endcsname}
		{\noexpand\orcidicon}}
}

\begin{document}
	
\title{Revisiting Cosmic Distance Duality with Megamasers and DESI DR2 Observations: Model Independent Constraints on Early–Late Calibration}

\author{Brijesh Kanodia\orcidA{}}
\email{brijeshk@iisc.ac.in}
\affiliation{Department of Physics, Indian Institute of Science,\\ C.\,V.\,Raman Road, Bengaluru 560012, India}
\affiliation{Centre for High Energy Physics, Indian Institute of Science,\\ C.\,V.\,Raman Road, Bangalore 560012, India}
	
\author{Ujjwal Upadhyay\orcidB{}}
\email{ujjwalu@iisc.ac.in}
\affiliation{Department of Physics, Indian Institute of Science,\\ C.\,V.\,Raman Road, Bengaluru 560012, India}
\affiliation{Astronomy \& Astrophysics Group, Raman Research Institute,\\ C.\,V.\,Raman Avenue, Bengaluru 560080, India}	
	
\author{Yashi Tiwari\orcidC{}}
\email{yashitiwari@iisc.ac.in}
\affiliation{Department of Physics, Indian Institute of Science,\\ C.\,V.\,Raman Road, Bengaluru 560012, India}
\affiliation{Astronomy \& Astrophysics Group, Raman Research Institute,\\ C.\,V.\,Raman Avenue, Bengaluru 560080, India}	
		
	%\date{\today}
	
\begin{abstract}
The Cosmic Distance Duality Relation (CDDR) connects the angular diameter distance ($d_A$) and the luminosity distance ($d_L$) at a given redshift. This fundamental relation holds in any metric theory of gravity, provided that photon number is conserved and light propagates along null geodesics. A deviation from this relation could indicate new physics beyond the standard cosmological model. In this work, we test the validity of the CDDR at very low redshifts (\( z < 0.04 \)) by combining $d_A$ from the Megamaser Cosmology Project with $d_L$ from the Pantheon+ sample of Type Ia Supernovae (SNIa). We further incorporate high-redshift Baryon Acoustic Oscillation (BAO)--based $d_A$ measurements from DESI DR2 in combination with SNIa data, highlighting the critical role of the $r_d-M_b$ (early–late) calibration in testing the CDDR using these two probes. Assuming CDDR holds, we perform a Bayesian analysis to derive model-independent constraints on the calibration parameters. Using only BAO and SNIa data, we observe a strong degeneracy between $r_d$ and $M_b$. However, the inclusion of calibration-free Megamaser measurements breaks this degeneracy, enabling independent constraints without relying on a specific cosmological model or distance-ladder techniques. Additionally, we present a forecast incorporating the expected precision from future Megamaser and SNIa observations, demonstrating their potential to significantly tighten constraints on early–late calibration parameters, under the assumption of validity of CDDR.
\end{abstract}

	
\maketitle

\section{Introduction} \label{sec:first}
The Cosmic Distance Duality Relation (CDDR) is a fundamental geometric relation in observational cosmology \cite{ 2007GReGr..39.1047E, Lima:2011ye}. At a given redshift \( z \), it is expressed as $d_L(z) = (1+z)^2 d_A(z)$, where \( d_L \) and \( d_A \) denote the luminosity distance and the angular diameter distance, respectively. This relation is a direct consequence of Etherington's reciprocity theorem \cite{etherington1933lx}, and holds under very general assumptions—namely, that light travels along null geodesics in a metric theory of gravity and that photon number is conserved. Because CDDR is independent of the specific cosmological model, it serves as a foundational consistency check in cosmological observations. Its validity is often assumed, sometimes implicitly, in a wide range of analyses \cite{2012A&A...538A.131H, 2010MNRAS.401.2331L, Cao:2011pb, Colaco:2025yhw, Colaco:2025aqp}. Therefore, any violation of this relation could have significant implications, potentially pointing to new or exotic physics, such as photon-axion conversion, violations of Lorentz invariance, or opacity effects \cite{Bassett:2003vu, Santos:2025gjf, Avgoustidis:2011aa, Ellis:2013cu, Giesel:2022org}. As a result, numerous efforts have been made to test the robustness of the CDDR using independent observational probes \cite{Uzan:2004my, DeBernardis:2006ii, Lazkoz:2007cc, Holanda_2010, Holanda_2011, Li:2011exa, Khedekar:2011gf, Gahlaut:2025lhv, Tang:2024zkc, Qi:2024acx, Yang:2024icv}.

\par Testing the validity of CDDR ideally requires measurements of \( d_A \) and \( d_L \) at the same redshifts. However, such direct tests are limited by the scarcity of reliable astrophysical objects for which both distances can be independently and accurately determined. A more practical and widely adopted approach involves combining different types of cosmological probes: specifically, luminosity distance $d_L$ from standard candles such as Type Ia Supernovae (SNIa), and angular diameter distance $d_A$ from standard rulers such as Baryon Acoustic Oscillations (BAO). Because these observational datasets typically differ in their redshift coverage, various statistical techniques, particularly interpolation or strategic binning of the $d_L$ measurements, have been employed to facilitate meaningful comparisons across redshift ranges \cite{2013MNRAS.436.1017L, Mukherjee:2021kcu}. 
\par Distance measurements in cosmology are inherently challenging and rely heavily on calibration. For instance, in the case of SNIa, determining \( d_L \) requires accurate knowledge of the absolute magnitude \( M_b \), which is sensitive to the underlying astrophysics and the calibration anchors used \cite{Riess:2021jrx, Camarena:2021jlr}. Similarly, inferring \( d_A \) from BAO observations depends on calibrating the sound horizon at the baryon drag epoch, \( r_d \), which in turn is governed by the physics of the early universe \cite{Planck:2018vyg, Bernal:2016gxb}. 
Consequently, although the CDDR is theoretically independent of the cosmological model, practical tests based on probes like SNIa and BAO involve model-dependent calibrations, making such tests only approximately model-independent. In fact, inconsistent choices of calibrations such as $r_d$ and $M_b$ can induce apparent violations of the CDDR. To avoid this, some studies have tested the CDDR using methods that eliminate the dependence on calibration parameters, finding results that remain consistent with the CDDR \cite{Tonghua:2023hdz, Yang:2025qdg, Wang:2024rxm}.
\par With the recent data release from the Dark Energy Spectroscopic Instrument (DESI)\cite{DESI:2024mwx, DESI:2025zgx}, providing increasingly precise measurements of the expansion history through BAO, several studies have revisited tests of the CDDR \cite{Yang:2025qdg, Wang:2025gus, Afroz:2025iwo, Zhang:2025qbs}. While most of these analyses find CDDR to be consistent within $2\sigma$, some report mild deviations at specific redshifts. For instance, Ref.~\cite{Afroz:2025iwo} identifies a deviation from CDDR, but interprets it as an inconsistency between SNIa and BAO datasets, rather than an actual breakdown of the relation itself. Implications of such apparent violations of CDDR have also been studied in the context of cosmological tensions and in assessing the robustness of the claim for dynamical dark energy suggested by DESI DR2 observations \cite{Afroz:2025iwo, Teixeira:2025czm, Alfano:2025gie, Dhawan:2025mer, Jesus:2024nrl}. However, as noted earlier, a critical consideration in testing CDDR—particularly when using BAO and SNIa data—is the role of the calibration parameters $r_d$ and $M_b$. This issue is also closely related to the Hubble tension, often referred to as the `cosmic calibration tension', as discussed in detail in Ref.~\cite{Poulin:2024ken}. These considerations also underscore the importance of calibration-independent, geometric distance probes for rigorously testing the validity of CDDR and for strengthening the robustness of cosmological inferences based on it.
\par The aim of this work is threefold. First, we test the validity of the CDDR at very low redshifts ($z < 0.04$) using angular diameter distance measurements from the Megamaser Cosmology Project (MCP) \cite{Pesce:2020xfe, 2020ApJ...890..118P}. These measurements are free from external calibration and serve as a robust, model-independent distance probe. When combined with luminosity distance data from SNIa, they enable a low-redshift test of the CDDR that has not been previously explored. Second, we revisit the standard approach of testing the CDDR using BAO and SNIa data, with particular focus on the influence of the calibration parameters $r_d$ and $M_b$. We demonstrate that the conclusions of such tests are highly sensitive to the choice of these parameters, underscoring the importance of calibration in CDDR analyses. In the third part of our analysis, we perform a detailed MCMC study using current data, assuming the validity of the CDDR to derive model-independent constraints on the early–late calibration parameters \( r_d \) and \( M_b \). We find that when only BAO and SNIa data are used, a strong degeneracy exists between these parameters. However, this degeneracy can be substantially lifted by incorporating the calibration-free Megamaser data. The resulting combined analysis, based solely on the assumption of CDDR validity, yields constraints on \( r_d \) and \( M_b \) that are independent of any cosmological model. In addition, we present a forecast based on expected improvements in future SNIa and Megamaser observations, illustrating the potential for significantly tighter constraints on these key calibration parameters.

\par This paper is organized as follows. In Section~\ref{sec:second}, we describe the datasets employed in this work along with the methodology used for the analysis. In Section~\ref{sec:third} we present the main results obtained using this methodology. Finally, we summarize and discuss the implications of our findings in Section~\ref{sec:fourth}.
\section{Data and Methodology} \label{sec:second}
In this section, we describe the datasets used in the analysis and provide a detailed outline of the methodology adopted.
\vspace{-0.3in}
\subsection{Data}
\label{sec: data}
The datasets used in this work are as follows,
\begin{itemize}
    \item {\bf Pantheon+:} The dataset comprises 1701 spectroscopically confirmed SNIa, providing their redshifts and apparent magnitudes along with the full covariance matrix that includes both statistical and systematic uncertainties. We use this uncalibrated data, which spans a redshift range of approximately $z\in(0.001,2.26]$ \cite{Scolnic:2021amr}.
    
    \item {\bf DESI DR2:} We utilize BAO measurements, specifically \( D_M / r_d \) along with their associated uncertainties, covering the redshift range \( z \sim 0.5 \) to \( 2.3 \), as reported in Table~IV of Ref.~\cite{DESI:2025zgx}. We use five of these measurements and exclude the data point at \( z \sim 2.3 \) due to the lack of sufficient SNIa data at that redshift.

     \item {\bf Megamaser:} We use maser-based measurements of the angular diameter distance and recession velocity for five galaxies, as reported by the MCP in Table~I of Ref.~\cite{Pesce:2020xfe}. The last data point in that table, corresponding to a very low redshift ($z\sim0.002$), is excluded from our analysis due to the limited number of SNIa data points available in that redshift regime.
\end{itemize}

\subsection{Testing CDDR with Observations}
\label{sec:methodology}
Testing the CDDR commonly involves using the ratio of the distances $d_L$ and $d_A$ to define a diagnostic function \cite{2013MNRAS.436.1017L},
\begin{equation}
\eta(z) \equiv \frac{d_L(z)}{(1+z)^2 d_A(z)}.
\label{e:eta}
\end{equation}
If the CDDR holds, then $\eta(z) = 1$ at all redshifts. In practice, however, measurements of $d_L$, $d_A$, and redshift $z$ are subject to statistical and systematic uncertainties.
A significant deviation of $\eta(z)$ from unity (typically exceeding $2\sigma$ uncertainity) can be interpreted as a potential violation of the relation. Assuming the errors in $d_L$, $d_A$, and $z$ are uncorrelated, the uncertainty in $\eta$ can be propagated as,
\begin{equation}
\sigma^2_{\eta} = \eta^2 \left[
\left( \frac{\sigma_{d_L}}{d_L} \right)^2 +
\left( \frac{\sigma_{d_A}}{d_A} \right)^2 +
\left( \frac{2 \sigma_z}{1+z} \right)^2
\right].
\label{e:eta_error}
\end{equation}
\subsubsection{Luminosity distance from SNIa} The luminosity distance $d_L(z)$ in megaparsecs derived from the apparent magnitude measurements $m(z)$ from Pantheon+ SNIa is given by
\begin{equation}
    d_L(z)=10^{(m(z)-M_b-25)/5}, 
    \label{e:DL}
\end{equation}
where $M_b$ denotes the absolute magnitude of the supernovae, serving as a calibration parameter. The Pantheon+ dataset includes a full covariance matrix for $m(z)$, accounting for both statistical and systematic uncertainties. This covariance can be propagated to the luminosity distance using
\begin{equation}
\mathrm{Cov}_{d_L} = \mathbf{J} \, \mathrm{Cov}_{m_B} \, \mathbf{J}^\top, \quad
\mathbf{J} = \mathrm{diag}\left( \frac{\ln 10}{5} d_L \right),
\label{eq:Jacobian}
\end{equation}
where $\mathbf{J}$ is the Jacobian matrix of the transformation. The uncertainties $\sigma_{d_L}$ are then extracted from the diagonal elements of the resulting covariance matrix. Redshift measurements in the Pantheon+ compilation are spectroscopic and thus have negligible uncertainties.
\subsubsection{Angular diameter distance from BAO} The angular diameter distance $d_A(z)$ used in this work is derived from BAO measurements and megamaser observations. BAO data provide the angular size of the sound horizon at the baryon drag epoch, defined as
\begin{equation}
    \theta_{\text{BAO}}(z)=\frac{r_d}{d_M(z)}=\frac{r_d}{d_A(z) (1+z)},
    \label{e:BAO-data}
\end{equation}
where $d_M$ is the comoving distance given by,
\begin{equation}
    d_M(z)=\int_0^z \frac{dz'}{H(z')},
    \label{e:DM}
\end{equation}
in a flat universe and related to the angular diameter distance by $d_A(z) = d_M(z)/(1+z)$. Thus, the angular diameter distance can be expressed as
\begin{equation}
d_A(z) =  \frac{r_d}{(1+z)\theta_{\text{BAO}}(z)},
\label{e:DA}
\end{equation}
where $r_d$ is the sound horizon at the drag epoch, serving as another calibration parameter. The uncertainty in $d_A$ is then propagated as
\begin{equation}
\sigma_{d_A}^2 = \frac{r_d^2}{(1+z)^2 \theta^2_{\rm BAO}} \sigma_{\theta_{\text{BAO}}}^2,
\label{e:sigma_DA}
\end{equation}
assuming the uncertainty in redshift is negligible due to the spectroscopic precision of the DESI BAO measurements. The calibration parameters $M_b$ and $r_d$ are determined from distance ladder methods and cosmological modeling, each carrying its own uncertainties. However, these uncertainties are relatively small compared to those in the measurements of $m(z)$ and $\theta_{\text{BAO}}(z)$, and thus have only a minor impact on the overall uncertainty in $\eta(z)$. For this reason, we neglect them in the present analysis. Nevertheless, we emphasize that the inferred $\eta(z)$ remains sensitive to the assumed mean values of these calibration parameters. Different choices arising, for example, from cosmological model dependence in $r_d$ or from late-time calibration systematics in $M_B$, can lead to noticeable shifts in $\eta(z)$ even when the quoted uncertainties on these parameters are small.
\subsubsection{Angular diameter distance from Megamasers} Water megamasers are powerful maser emissions originating from the thin, warped molecular disks orbiting supermassive black holes (SMBHs) in active galactic nuclei (AGNs). They emit at microwave frequencies corresponding to the rotational transitions of water vapour. These systems enable precise, one-step geometric distance measurements to their host galaxies, independent of the cosmic distance ladder. This is achieved by combining Very Long Baseline Interferometry (VLBI) measurements of angular positions with spectroscopic data on velocities and centripetal accelerations of maser spots in Keplerian motion \cite{Pesce:2020xfe, 2020ApJ...890..118P}. The analysis provides angular diameter distance measurements, $d_A(z)$, to host galaxies at extragalactic scales, along with their recession velocities, $v(z)$—and hence redshifts $z$—as well as the associated statistical uncertainties. These uncertainties can be appropriately propagated into the estimation of $\eta(z)$.
However, since these measurements are at very low redshifts, particular care is needed when interpreting the observed recession velocity as a proxy for cosmological redshift. This is because the observed velocity, $v$, includes contributions from both the Hubble flow and the galaxy's peculiar motion, $v_{\rm pec}$, such that $v=v_{\text{pec}}+cz$. This introduces an additional systematic uncertainty that becomes especially important when performing parameter estimation within a given cosmological framework. However,
for the purpose of testing the validity of the CDDR in a model-independent manner, uncertainties in redshift can be safely neglected, as they are significantly smaller than the uncertainties associated with the measurements of \( d_L \) and \( d_A \). In the next section, we present a more refined treatment of peculiar velocity effects, where these contributions are explicitly incorporated into the velocity uncertainty when using megamaser data for parameter inference within a specified cosmological framework. In this work, we use the $d_A$ measurements of the five Megamaser hosting galaxies along with their recession velocities as reported by the MCP in Table 1 of \cite{Pesce:2020xfe}. 
\par The angular diameter distance measurements from DESI BAO observations primarily cover the high-redshift range, $z \sim 0.5$–2.3. In contrast, the megamaser measurements from the MCP are confined to very low redshifts, $z < 0.04$, offering a unique opportunity to test the validity of the CDDR in the nearby Universe. The luminosity distances from the Pantheon+ SNIa dataset span a broad redshift range, overlapping with both the BAO and megamaser $d_A$ measurements, thus enabling a consistent comparison across cosmic epochs. To enable a direct test of the CDDR at the redshifts corresponding to BAO and megamaser observations, we reconstruct $d_L(z)$ using a model-independent Gaussian Process Regression (GPR) applied to the Pantheon+ SNIa data. The details of the reconstruction method are provided in Appendix \ref{sec:appendix}. Finally, the framework outlined above enables a consistent and approximately model-independent estimation of $\eta(z)$ and its associated uncertainty, providing a robust test of the distance duality relation.
\subsection{Calibration Degeneracy in CDDR Test}
\label{sec: calibration-degeneracy}
As discussed in the previous section, testing the CDDR using a combination of BAO and SNIa data requires specifying the calibration parameters $r_d$ and $M_b$, which are associated with the BAO and SNIa datasets, respectively. In Sec. \ref{sec: 3A}, we examine how different choices of these parameters impact the evaluation of $\eta(z)$ and influence the interpretation of CDDR validity.
\par In this section, we take a complementary approach by leveraging the CDDR as a consistency check between independent datasets, allowing us to constrain the calibration parameters without assuming a specific cosmological model. In particular, we focus on the degeneracy between $r_d$ and $M_b$, highlighting what we believe to be a novel application of the CDDR framework.
Assuming the CDDR holds exactly, i.e., $\eta(z) = 1$, and substituting Eqs. (\ref{e:DL}) and (\ref{e:DA}) into Eq. (\ref{e:eta}), we arrive at the following relation,
\begin{equation}
    M_b + 5 \log r_d = m(z) + 5 \log \left[ \frac{\theta_{\mathrm{BAO}}(z)}{(1+z)}  \right]-25
\end{equation}
which captures the degeneracy between the two calibration parameters. The left-hand side of the equation represents a specific combination of the calibration parameters $r_d$ and $M_b$, while the right-hand side is entirely determined by observational data from SNIa and BAO measurements. Since both $r_d$ and $M_b$ are constants, the equation can be recast as
\begin{equation}
    A=\mathcal{F}\Big(m(z),\theta_{\text{BAO}}(z)\Big)
\end{equation}
where the constant A is given as,
\begin{equation}
    A \equiv M_b + 5\log r_d
    \label{e:rd_Mb_rel}
\end{equation}
and the function $\mathcal{F}$ is given by
\begin{align}
    \mathcal{F}& \equiv  m(z) + 5 \log \left[ \frac{\theta_{\mathrm{BAO}}(z)}{(1+z)} \right]-25
\end{align}
Thus, any given pair of BAO–SNIa observations constrains only the combination $M_b + 5 \log r_d$, as expressed in Eq. (\ref{e:rd_Mb_rel}). In other words, for a fixed dataset, multiple combinations of $r_d$ and $M_b$ can satisfy the CDDR, highlighting a degeneracy between these calibration parameters.
Additionally, the degeneracy between $r_d$ and $M_b$, assuming the validity of the CDDR, can be robustly explored using current BAO and SNIa data through standard Markov Chain Monte Carlo (MCMC) techniques. In the following section, we outline the MCMC methodology for the same and show how incorporating calibration-free megamaser data helps to break this degeneracy, allowing for model-independent constraints on the calibration parameters.
\subsection{Bayesian Constraints and Forecast}
\label{sec: method_bayesian}
In this section, we describe the Bayesian framework employed to incorporate the CDDR in order to constrain the early–late calibration parameters. Using SNIa, BAO, and Megamaser data as outlined in Sec.~\ref{sec: data}, we perform an MCMC analysis to derive the corresponding constraints. Under the assumption that the CDDR holds, the likelihood function is given by,
\begin{equation}
    \mathcal{L}_{\eta}(r_d, M_b) \propto \prod_i \exp\left[-\frac{(\eta_i - \eta_0)^2}{2\sigma_{\eta_i}^2}\right],
    \label{e:likelihood}
\end{equation}  
where we set \(\eta_0 = 1\), which corresponds to the theoretical expectation from the CDDR. The \(\eta_i\) are the distance duality ratios defined in Eq.~(\ref{e:eta}) at redshifts \(z_i\), and \(\sigma_{\eta_i}\) are their corresponding uncertainties, calculated using Eq.~(\ref{e:eta_error}). The index \(i\) runs over redshifts where BAO or Megamaser data are available, and where \(\eta_i\) is computed using the reconstructed \(d_L(z)\) from SNIa observations. Since the computation of \(\eta_i\) involves quantities derived from data that depend on the calibration parameters, the likelihood in Eq.~(\ref{e:likelihood}) becomes a function of both \(r_d\) and \(M_b\). Specifically, the luminosity distance inferred from SNIa depends on \(M_b\) as shown in Eq.~(\ref{e:DL}), while the angular diameter distances from BAO depend on \(r_d\) via Eq.~(\ref{e:DA}). In contrast, Megamaser-based \(d_A\) measurements are calibration-independent. 
\par However, when using Megamaser data—which provide recession velocities of maser-host galaxies—the inference of cosmological redshift from these velocities requires special care, as mentioned in Sec.~\ref{sec:methodology}. This is particularly important at low redshifts, where peculiar motions can contribute significantly to the observed velocities. If not properly accounted for, these effects can bias cosmological inferences (for example, see Ref.~\cite{Upadhyay:2025oit} and references therein). To address this, we incorporate an additional likelihood term that corrects for peculiar velocities in the Megamaser redshift estimates, following the approach of Ref.~\cite{Pesce:2020xfe},
\begin{equation}
\mathcal{L}_{v_p}(\hat{z}) = \prod_i \frac{1}{\sqrt{2\pi(\sigma_{z,i}^2 + \sigma_{z,\mathrm{pec}}^2)}} 
\exp\left( -\frac{1}{2} \frac{(z_i - \hat{z}_i)^2}{\sigma_{z,i}^2 + \sigma_{z,\mathrm{pec}}^2} \right),
\end{equation}
where $z_i\approx v_i/c$ is the redshift derived from observed Megamaser recession velocity $v_i$, and \( \hat{z}_i \) is the contribution from peculiar velocities. These \( \hat{z}_i \) values are treated as nuisance parameters in our analysis. Here, \( \sigma_{z,i} \) is the statistical uncertainty in the redshift estimate, and \( \sigma_{z,\mathrm{pec}} \) captures the redshift dispersion due to peculiar velocities, which we take to be \( \sim 250\,\mathrm{km\,s^{-1}} \).
The total likelihood is then given by,
\begin{equation}
    \mathcal{L} \propto \mathcal{L}_{\eta} \times \mathcal{L}_{v_p},
\end{equation}
where \( \mathcal{L} = \mathcal{L}(r_d, M_b, \hat{z}_i) \), and the nuisance parameters \( \hat{z}_i \) are marginalised over to obtain constraints on the calibration parameters \( r_d \) and \( M_b \).
\par For the MCMC analysis, we implement the above likelihood function in a custom Python code and run 12 parallel chains, each with \(10^7\) steps. Convergence is monitored using the Gelman--Rubin diagnostic~\cite{Gelman:1992}, ensuring \( R - 1 < 0.01 \). The quantities \( \hat{z}_i \) are treated as nuisance parameters and are sampled five times more frequently than the calibration parameters \( r_d \) and \( M_b \) to enhance sampling efficiency in the marginalisation process. Uniform (flat) priors are adopted for all parameters, with the ranges: $r_d  \in [120, 155]$, $M_b \in [-20.0, -19.0]$, and $\hat{z}_i  \in [-0.003, 0.003]$, corresponding to peculiar velocities $v_p  \in [-1000, +1000]~\mathrm{km/s}$. 
\par For the forecast analysis, we aim to estimate the expected improvement in the uncertainties of \( r_d \) and \( M_b \) with future data. To this end, we adopt $\sigma_{m} = 0.02$, consistent with the projected precision from upcoming LSST observations~\cite{LSSTScienceBook2018}, and assume a $5\%$ statistical uncertainty for $d_A$ from Megamaser measurements~\cite{Shaw2024}.
 For the BAO data, we adopt the current high-precision constraints from DESI DR2~\cite{DESI:2025zgx}. It is important to note that the primary focus of this forecast is to assess the reduction in the uncertainties of \( r_d \) and \( M_b \) relative to present constraints. While the central values of these parameters may shift in either direction with future data, our analysis is directed toward quantifying the expected gains in precision.
\section{Results} \label{sec:third}
We now present the results of our analysis, structured into three parts following the methodology outlined in the previous section. 
\vspace{-0.1in}
\subsection{Low Redshift Test of CDDR with Megamasers and SNIa}
\label{sec: 3A}

\begin{figure}[b!]
    \centering
    \includegraphics[width=\linewidth]{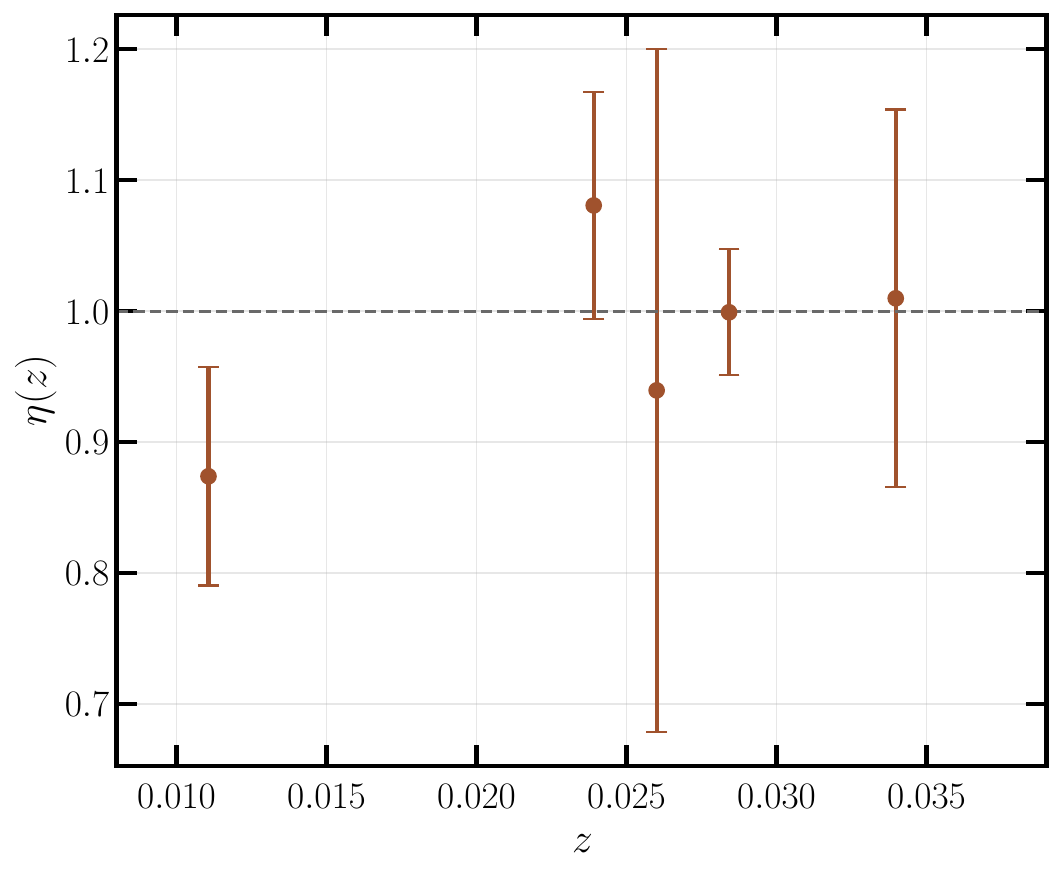}
    \caption{The distance duality ratio $\eta(z)$ obtained using reconstructed $d_L$ from SNIa and $d_A$ from Megamaser data \cite{Pesce:2020xfe}. Error bars represent the $1\sigma$ measurement uncertainties, incorporating contributions from both SNIa and Megamaser data. The ratio $\eta(z)$ is expected to be unity, indicated by the dashed line, if CDDR holds. 
    The supernova distances are computed using \( M_b = -19.25 \), though the resulting \( \eta(z) \) remains consistent with unity under small variations in \( M_b \).}
    \label{fig: megamaser-CDDR}
\end{figure}
In this subsection, we present the results of testing CDDR using megamaser data. Figure~\ref{fig: megamaser-CDDR} shows the ratio \( \eta \), constructed using \( d_A \) from megamasers and \( d_L \) from supernovae. Megamaser $d_A$ measurements allow testing the CDDR at very low redshifts (\( z < 0.04 \)), a regime inaccessible to BAO, quasars, or galaxy clusters, which primarily cover higher redshifts (\( z > 0.1 \)). As shown in the figure, the results are consistent with CDDR within \( 1\sigma \) uncertainties. From the combined data points, we find the mean value, \( \bar\eta = 0.9805 \pm 0.0697 \), in excellent agreement with the theoretical expectation of unity. The relatively large error bars are primarily due to the current \( \sim 15\% \) uncertainty in megamaser-based \( d_A \) measurements, which is expected to improve to below \( 5\% \) with future observations \cite{Shaw2024}. Despite the present limitations, the calibration-free nature of megamasers makes them a valuable probe, offering promising applications such as independent measurements of the Hubble constant \cite{Pesce:2020xfe} and robust tests of the distance duality relation in the local universe.

\subsection{Calibration Dependence of CDDR with \\BAO and SNIa}
\begin{figure}[tbh!]
    \centering
    \includegraphics[width=\linewidth]{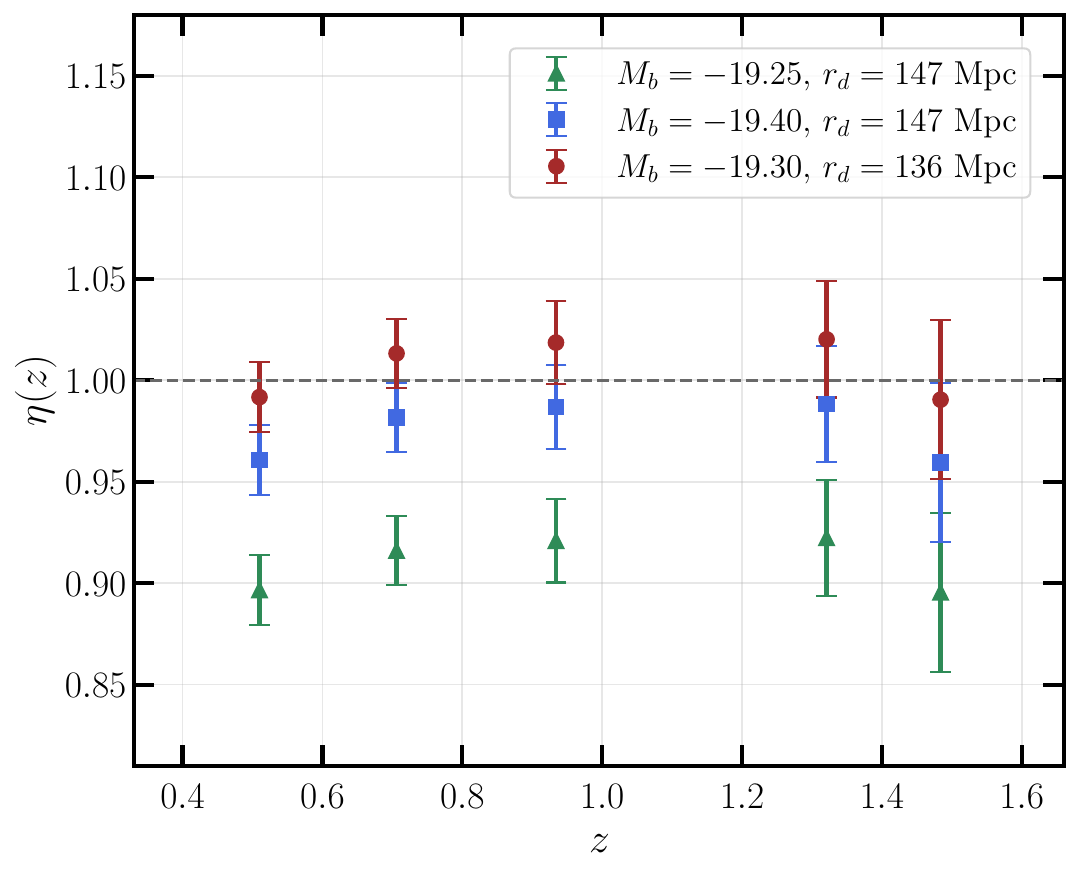}
    \caption{The distance duality ratio $\eta(z)$ obtained using reconstructed $d_L$ from SNIa and $d_A$ from DESI BAO data \cite{DESI:2025zgx}, for different choices of calibration parameters. Error bars represent the $1\sigma$ measurement uncertainties, incorporating contributions from both SNIa and BAO data. The ratio $\eta(z)$ is expected to be unity, indicated by the dashed line, if CDDR holds.}
    \label{fig:eta_bao}
\end{figure}

In this subsection, we explore how different choices of the calibration parameters \( r_d \) and \( M_b \) impact the validity of the CDDR when combining BAO and SNIa data. Figure~\ref{fig:eta_bao} shows the distance duality ratio \( \eta(z) \) calculated at the redshift of BAO data for various combinations of \( r_d \) and \( M_b \), highlighting their influence on the inferred values of \( \eta(z) \). Before examining the implications of these choices, we first outline the motivation behind selecting specific values for the calibration parameters.
\par The low-redshift distance ladder approach, as reported by the SH0ES collaboration~\cite{Riess:2021jrx, Camarena:2021jlr}, measures the absolute magnitude of SNIa to be \( M_b = -19.25 \pm 0.03 \), leading to a Hubble constant of \( H_0 = 73.2 \pm 1.3 \,\mathrm{km\,s^{-1}\,Mpc^{-1}} \). This estimate is in strong tension ($>5\sigma$) with the value inferred from early-universe observations, such as the CMB and BAO, which prefer \( M_b \approx -19.4 \) and \( H_0 = 67.4 \pm 0.5 \,\mathrm{km\,s^{-1}\,Mpc^{-1}} \) under the standard \( \Lambda \)CDM model~\cite{Planck:2018vyg}. This discrepancy is a restatement of the widely discussed Hubble tension--a persistent and statistically significant disagreement between early- and late-time determinations of \( H_0 \)~\cite{Abdalla:2022yfr, DiValentino:2021izs, Schoneberg:2021qvd}. The inference of \( H_0 \) from early-universe data is sensitive to the calibration of the standard ruler \( r_d \), which within the \( \Lambda \)CDM framework and Planck CMB data, is tightly constrained to \( r_d = 147.1 \pm 0.3 \) Mpc~\cite{Planck:2018vyg}. Reconciling the high local value of \( H_0 \) with early-universe constraints requires reducing the value of \( r_d \) to approximately \( 140 \) Mpc or below~\cite{Bernal:2016gxb, Poulin:2018cxd, Knox:2019rjx}, thus allowing for a higher inferred expansion rate consistent with distance-ladder measurements.

In Figure~\ref{fig:eta_bao}, the green data points corresponding to the combination \( r_d = 147 \) Mpc and \( M_b = -19.25 \) exhibits a noticeable deviation from the expected value \( \eta(z) = 1 \), indicating a violation of the CDDR. This case effectively captures the essence of the Hubble tension, reflecting the discrepancy between the early- and late-universe calibration parameters \( r_d \) and \( M_b \). In contrast, the blue points—representing the combination \( r_d = 147 \) Mpc and \( M_b = -19.40 \), consistent with \( \Lambda \)CDM and early-universe data—show excellent agreement with CDDR. Similarly, the red points, corresponding to \( r_d = 136 \) Mpc and \( M_b = -19.30 \), motivated by early-universe solutions to the Hubble tension, also satisfy the CDDR well. These cases collectively illustrate the sensitivity of the CDDR test to the choice of calibration parameters, with similar conclusions also drawn in Ref. \cite{Teixeira:2025czm}. In other words, the apparent validity or violation of the distance duality relation strongly depends on the assumed values of \( r_d \) and \( M_b \) when working with BAO and SNIa data. This interplay is examined in greater detail in the following section.
These results further underscore the limitations of testing the CDDR using calibration-dependent distance probes—particularly dependence on parameters \( r_d \) and \( M_b \)—whose careful selection is crucial in the current era of precision cosmology and growing cosmological tensions.
\subsection{Constraints on Early-Late Calibration from CDDR and Forecast}
In this subsection, assuming the validity of the CDDR, we present cosmology-independent constraints on the calibration parameters $r_d$ and $M_b$. The analysis is performed using the MCMC framework described in Sec.~\ref {sec: method_bayesian}, and combining current SNIa, BAO, and Megamaser data mentioned in Sec.~\ref {sec: data}. We further include a Bayesian forecast based on projected uncertainties from future SNIa and Megamaser measurements, illustrating the anticipated improvement in precision.
\par
\begin{figure*}[t]
    \centering
    % Left figure
    \begin{minipage}[t]{0.48\textwidth}
        \centering
        \includegraphics[scale=0.80]{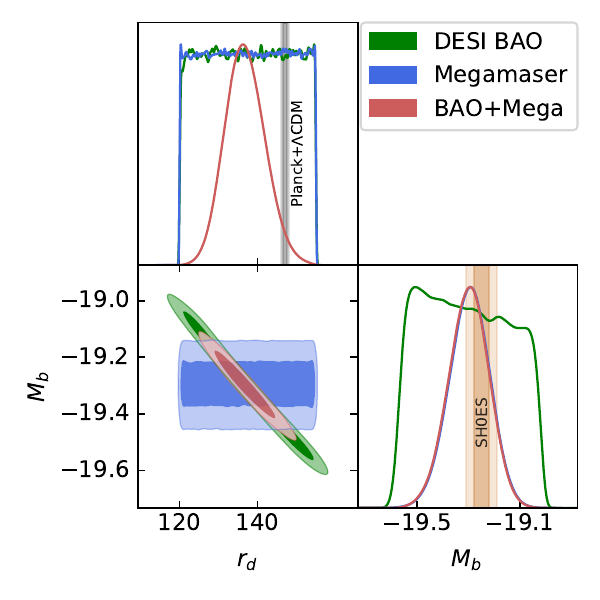}
        \vspace{-0.5em}
        %\caption*{(a)}
    \end{minipage}
    \hfill
    % Right figure
    \begin{minipage}[t]{0.48\textwidth}
        \centering
        \includegraphics[scale=0.80]{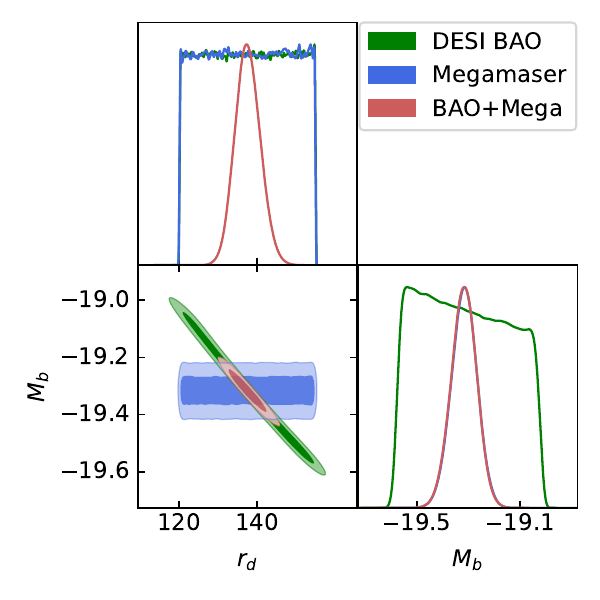}
        \vspace{-0.5em}
        %\caption*{(b)}
    \end{minipage}

    \caption{Constraints (left) and forecast (right) for the calibration parameters assuming the validity of CDDR. Colors for 1D and 2D posteriors represent the datasets used in combination with Pantheon+ SNIa. The vertical bands in the left panel show constraints on $r_d$ and $M_b$ from Planck18+$\Lambda$CDM and SH0ES collaboration, respectively.  The inclusion of Megamaser data, alongside DESI BAO and Pantheon+, breaks the $r_d-M_b$ degeneracy.}
    \label{fig:con_and_for}
\end{figure*}

 Figure~\ref{fig:con_and_for} displays the posterior distributions for \( r_d \) and \( M_b \), along with the corresponding \(1\sigma\) and \(2\sigma\) credible regions. The left panel shows current constraints from existing datasets, while the right panel presents forecasted results, highlighting the expected reduction in uncertainties in the measurement of these parameters. The three colors represent different choices for the angular diameter distance probes: BAO (calibrated via \( r_d \)), Megamasers (calibration-independent), and their combination. In all cases, the luminosity distance is derived from SNIa data calibrated through \( M_b \).
When using BAO data (green), we observe a strong degeneracy in the \( r_d \)–\( M_b \) plane, indicating that the validity of the CDDR constrains only a combination of these two parameters, rather than each individually. This essentially reflects the degeneracy relation discussed in Sec.~\ref{sec: calibration-degeneracy}.
In contrast, when using calibration-independent Megamaser data (blue), the CDDR primarily constrains \( M_b \). Notably, assuming the validity of the CDDR across cosmic epochs enables a combined analysis with BAO and Megamaser data. Since the calibration parameters are redshift independent, the low redshift constraint on $M_b$ from Megamasers is sufficient to break the calibration degeneracy and yields independent constraints on both \( r_d \) and \( M_b \), as shown in the combined case (red). The joint constraints on the early–late calibration parameters, obtained by assuming the validity of the CDDR with current data, are given by,
\begin{align}
    r_d &= 137.5 \pm 5 \, \text{Mpc}, \nonumber \\
    M_b &= -19.3 \pm 0.08.
\end{align}
In Figure~\ref{fig:con_and_for}, we also highlight the \(2\sigma\) uncertainity bands for \( r_d \) and \( M_b \), corresponding to Planck18 CMB measurements within the \( \Lambda \)CDM framework \cite{Planck:2018vyg} and SH0ES distance-ladder results \cite{Riess:2021jrx}, respectively. Given the current observational precision of the data used in our analysis, the resulting constraints on the calibration parameters remain relatively weak. Nevertheless, they are fairly consistent within \(2\sigma\) with the predictions of the standard \( \Lambda \)CDM model; however, the mean value of \( r_d \) tends to peak toward a reduced sound horizon. We also note that the effect of peculiar velocity corrections for Megamasers is negligible and does not impact the inferred constraints, primarily due to the comparatively larger uncertainties in their distance measurements.
\par We now briefly discuss the forecast results shown in the right panel of Figure~\ref{fig:con_and_for}. While current uncertainties on \( r_d \) and \( M_b \) remain relatively large, future observations are expected to significantly tighten these bounds, as illustrated in the figure. In particular, based on the anticipated precision improvements in SNIa and Megamaser observations from LSST and MCP, respectively (as outlined in Sec.~\ref{sec: method_bayesian}), we find projected uncertainties, \( \sigma_{M_b} = 0.04 \) and \( \sigma_{r_d} = 2.5 \) Mpc, representing nearly a factor-of-two improvement over current constraints. These forecasts are particularly relevant in light of upcoming data, which will also determine the direction of shift in the mean values of \( r_d \) and \( M_b \), potentially offering valuable insights into the ongoing Hubble tension.
\par The above analysis demonstrates the use of the CDDR to obtain robust constraints on the cosmic calibration parameters, independently of specific cosmological models or distance ladder techniques, a similar approach also explored in Ref.~\cite{Gomez-Valent:2021hda}. This method relies solely on the assumption of the validity of the distance duality relation, which is strongly supported by numerous studies using a broad range of observational probes. Within this framework, the analysis not only complements existing approaches for constraining calibration parameters but can also provide a valuable perspective on the persistent and yet unresolved Hubble tension.

\section{Conclusion and Discussion}\label{sec:fourth}
The field of cosmology has progressed into the precision era, driven by a wealth of ongoing and upcoming observational surveys that probe the universe with unprecedented depth and accuracy. A central objective is to understand the nature of dark energy by precisely tracing the late-time expansion history, using probes such as BAO and SNIa. These datasets are often combined to leverage their complementary strengths and to place tighter constraints on cosmological parameters. However, the resulting inferences are also sensitive to the choice of calibration parameters associated with these probes, which play a critical role in shaping the conclusions. Another related concern is the Hubble tension, which remains an unresolved discrepancy~\cite{Abdalla:2022yfr, DiValentino:2021izs, Schoneberg:2021qvd} and is often recasted as an \( M_b \) tension \cite{Camarena:2021jlr} or cosmic calibration tension \cite{Poulin:2024ken}. The calibration parameter—whether \( r_d \) in early-universe probes or \( M_b \) in distance-ladder methods—directly influences the inferred value of \( H_0 \), and lies at the heart of ongoing efforts to resolve this tension~\cite{Chen:2024gnu, Poulin:2024ken}.
\par The Cosmic Distance Duality Relation is a fundamental relation that provides a powerful, theory-agnostic consistency check of our understanding of cosmic geometry and photon propagation. Any violation of this relation could signal new physics, such as photon number non-conservation, deviations from metric gravity, or exotic photon interactions, with profound implications for cosmology~\cite{Bassett:2003vu, Santos:2025gjf, Avgoustidis:2011aa, Ellis:2013cu, Giesel:2022org}. Numerous studies have tested the CDDR using distance measurements from a variety of observational probes, consistently finding agreement within $2\sigma$, with small deviations typically attributed to tensions or systematics in the underlying datasets \cite{Uzan:2004my, DeBernardis:2006ii, Lazkoz:2007cc, Holanda_2010, Holanda_2011, Li:2011exa, Khedekar:2011gf, Gahlaut:2025lhv, Tang:2024zkc, Qi:2024acx, Yang:2024icv}.
\par An alternative and equally important perspective on the CDDR is its role as a fundamental consistency check for distance measurements derived from diverse observational probes such as BAO and SNIa. Since both rely on calibration parameters to infer geometric distances, the assumption of CDDR validity enables model-independent constraints on these parameters. In particular, combining BAO and SNIa data within this framework allows one to test the mutual consistency of early- and late-time distance indicators, offering insights into the interplay between the cosmic calibration parameters.
\par In this work, we revisit the CDDR, with a particular emphasis on understanding the impact of the calibration parameters $r_d$  and $ M_b$ on tests of its validity. We demonstrate that CDDR tests based on SNIa and BAO measurements are highly sensitive to the choice of these calibration parameters. Thereby, drawing conclusions based on a fixed choice of calibration can lead to apparent violations of CDDR, which may in turn affect broader cosmological inferences. We investigate the degeneracy between $r_d$ and $M_b$ that arises from enforcing the validity of CDDR. Using the latest DESI DR2 BAO measurements in combination with the Pantheon+ SNIa sample, we derive model-independent constraints in the $r_d-M_b$ plane.
\par A novel aspect of this work is the use of calibration-free megamaser angular diameter distance measurements from the MCP at very low redshifts ($z<0.04$) to test the CDDR---a regime otherwise inaccessible to BAO or quasar-based $d_A$ measurements. Another key advantage of using Megamaser-based $d_A$ measurements is that, being calibration-free, they provide an independent way to constrain the absolute magnitude \( M_b \) of SNIa under the assumption of CDDR. However, these measurements are also subject to astrophysical uncertainties—such as the inclination of the maser disk, the mass of the central black hole, and related factors—that can bias the inferred $d_A$ values and may require improved modelling to properly account for such systematics~\cite{2020ApJ...890..118P,2011ApJ...727...20K}. When Megamaser-based $d_A$ measurements are combined with BAO-based $d_A$ measurements, the degeneracy in the $r_d$–$M_b$ plane can be broken, demonstrating the utility of CDDR as a tool for obtaining cosmology-independent constraints on early–late calibrations. Using current data, we find $r_d = 137.5 \pm 5.0$ Mpc and $M_b = -19.3 \pm 0.08$, both broadly consistent with CMB-based predictions under the standard $\Lambda$CDM model. Although the calibration uncertainties are larger than those inferred from Planck CMB measurements, they are derived without assuming any specific cosmological model. We further perform a Bayesian forecast incorporating expected improvements in future SNIa and Megamaser data, which is found to substantially reduce the uncertainties on these parameters.
\par The $H_0$ (or $M_b$) tension presents a significant challenge, highlighting the need for independent probes that allow calibration-free, one-step distance measurements of $H_0$. Although current megamaser observations are limited to a small number of suitable maser systems, future surveys are expected to expand the sample and reduce distance uncertainties, potentially bringing the precision on $H_0$ below 2\% \cite{Pesce:2020xfe, Shaw2024}. Meanwhile, the LSST survey at the Vera C. Rubin Observatory will vastly improve supernova statistics and reduce observational systematics \cite{LSSTScienceBook2018}. 
\par The CDDR offers a powerful model-independent consistency test to constrain key calibration parameters such as $r_d$ and $M_b$, that can shed deeper insights on the origin of current cosmological tensions. As such, it serves not only as a diagnostic tool but also as a pathway toward uncovering possible new physics or unaccounted systematics in the standard cosmological framework.

%%%%%%%%%%%%%%%%%%%%%%%%%%%%%%%%%%%%%%%
\section*{Acknowledgments} \label{sec:acknowledgments}
We thank Shiv Sethi, Rajeev Kumar Jain, Jishnu P. Sai, and Suvodip Mukherjee for their helpful comments and feedback. We are also thankful to Subhadip Bouri for his suggestions on the manuscript. BK acknowledges Shiuli Chatterjee for discussion on the draft. YT acknowledges the Raman Research Institute for hosting her as a visiting fellow during this project. 
\appendix
\section{Methodology for Luminosity Distance Reconstruction}
\label{sec:appendix}
To evaluate the distance duality ratio \( \eta \) at the redshifts corresponding to the BAO and Megamaser data points, we reconstruct the luminosity distance \( d_L \) using the Pantheon+ SNIa dataset \cite{Scolnic:2021amr}. For this purpose, we employ Gaussian Process Regression (GPR)~\cite{2020arXiv200910862W,10.5555/1162264,b60dec2e1b6c416387f33e9de784f573}, a non-parametric Bayesian technique that enables a smooth and model-independent reconstruction of \( d_L(z) \). Unlike conventional interpolation methods such as polynomial or spline fitting, GPR inherently accounts for observational uncertainties and correlations, making it especially well-suited for estimating \( \eta(z) \) and ensuring accurate error propagation throughout the analysis.
\begin{figure}[h!]
    \centering
    \includegraphics[width=\linewidth]{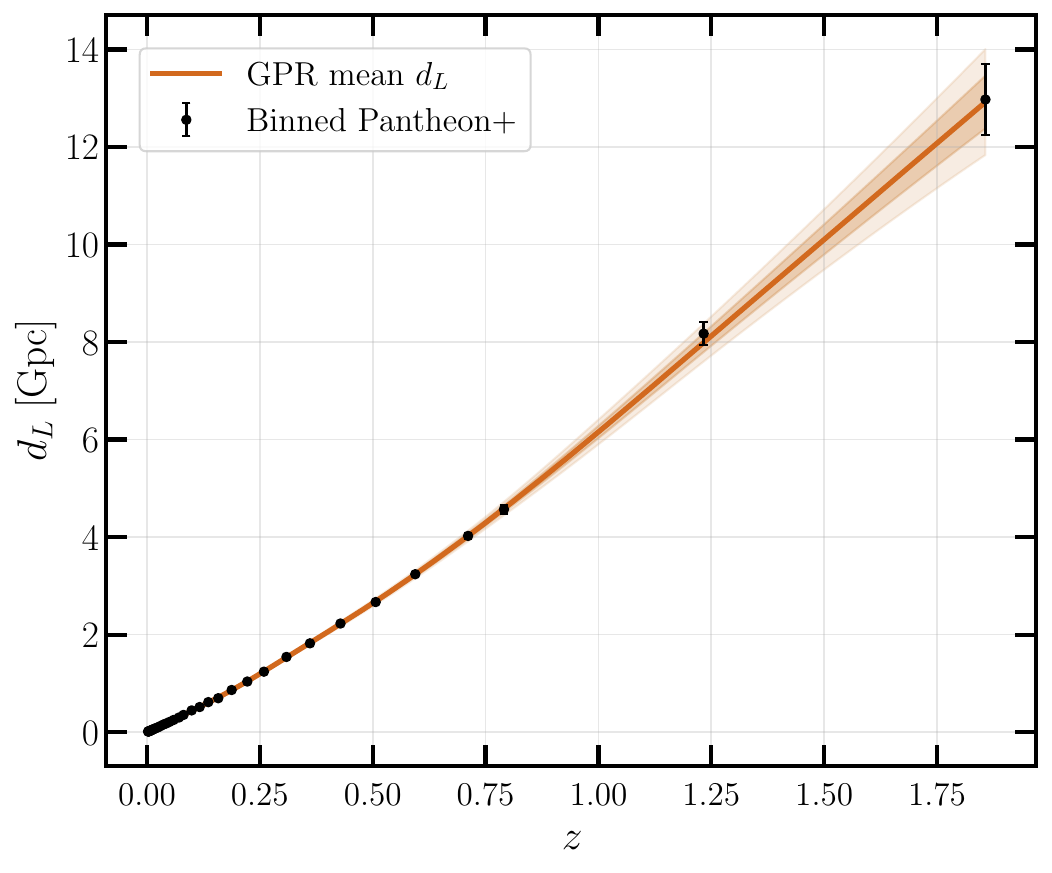}
    \caption{Gaussian Process reconstruction of the luminosity distance $d_L(z)$ using the Pantheon+ SNIa for $M_b=-19.25$. Black points represent binned data points, while the brown solid line indicates the GPR mean prediction. The shaded regions correspond to 1$\sigma$ and 2$\sigma$ uncertainty bands.}
    \label{fig: DL_gpr}
\end{figure}
\par The Pantheon+ sample contains 1701 SNIa, densely distributed at low redshifts and increasingly sparse at higher redshifts. To mitigate biases from this non-uniform redshift coverage, particularly at high \( z \), we adopt an adaptive binning strategy in magnitude–redshift space. The data are divided at a threshold of \( z = 0.9 \), below which logarithmic binning captures the fine structure of the densely sampled low-redshift region. While at $z>0.9$, coarser linear binning preserves statistical reliability in the sparsely populated high-redshift regime. Within each bin, we compute inverse-variance weighted averages of the observed apparent magnitudes \( m(z) \). These binned magnitudes are then converted to luminosity distances using the standard relation,
\begin{equation}
 d_L(z) = 10^{(m(z) - M_b - 25)/5},
\end{equation} 
where $M_b$ denotes the absolute magnitude of SNIa.  To eliminate the dependence of \( d_L(z) \) on \( M_b \), we define a rescaled luminosity distance,
\begin{equation}
\Tilde{d}_L(z)\equiv\frac{d_L(z)}{\lambda} = 10^{m(z)/5},
\end{equation} 
with a scaling factor $\lambda = 10^{-(M_b + 25)/5}$. The quantity $\Tilde{d}_L(z)$ depends solely on the observational data, which helps remove the dependence on the choice of calibration in the process of reconstruction. This, in turn, allows flexibility for subsequent MCMC analysis where $M_b$ is treated as a free parameter to obtain best-fit constraints. The rescaled luminosity distance $\Tilde{d}_L(z_i)$ is modelled using GPR,
\begin{equation} 
\Tilde{d}_L(z) \sim \mathcal{GP}(0, k(z,z')), 
\end{equation} 
where $k(z,z')$ is the covariance kernel. We work with the Matérn kernel~\cite{stein1999interpolation} with smoothness parameter $\nu = 3/2$ given by,
\begin{equation} 
k(z,z') = \sigma_f^2 \left(1+\frac{\sqrt{3}|z-z'|}{\ell}\right)\exp\left(-\frac{\sqrt{3}|z-z'|}{\ell}\right), 
\end{equation} 
with hyperparameters $\sigma_f$ representing the amplitude of fluctuations and $\ell$ defining the characteristic correlation scale.

Given observational data \{$z_i, \Tilde{d}_L(z_i)$\} with uncertainties derived from the Pantheon+ magnitude covariance matrix, the  posterior predictive distribution at an arbitrary redshift $z_*$ is Gaussian distribution with the mean and variance given by
\begin{align} 
\mu(z_*) &= \mathbf{k}*^\top (\mathbf{K} + \boldsymbol{\Sigma})^{-1}\mathbf{\Tilde{d}_L},\\
\sigma^2(z_*) &= k(z_*,z_*) - \mathbf{k}_*^\top(\mathbf{K}+\boldsymbol{\Sigma})^{-1}\mathbf{k}_*, 
\end{align} 
where $\mathbf{K}{ij}=k(z_i,z_j)$, $\mathbf{k}_*=[k(z_*,z_1),\ldots,k(z_*,z_n)]^\top$, and $\boldsymbol{\Sigma}=\mathrm{diag}[ \sigma_{\Tilde{d}_L,i}^2]$ represents observational uncertainties.

We adopt the optimal GPR hyperparameters \( (\sigma_f, \ell) \) and compute the weighted mean squared error at the training points, explicitly accounting for the non-uniform uncertainties across redshift. The rescaled luminosity distance \( \tilde{d}_L(z) \) is then converted to the physical luminosity distance \( d_L(z) \) by multiplying with the scaling factor \( \lambda \). Figure ~\ref{fig: DL_gpr} shows the reconstructed $d_L(z)$ for $M_b=-19.25$ corresponding to SH0ES calibration \cite{Riess:2021jrx}.
This combined strategy of optimized adaptive binning and Gaussian Process Regression offers a robust, model-independent reconstruction of the luminosity distance.

\bibliography{references.bib} 
\end{document}